\documentclass{IEEEtran}

\usepackage{cite}

\ifCLASSINFOpdf
  \usepackage[pdftex]{graphicx}
  \DeclareGraphicsExtensions{.pdf,.jpeg,.png}
\else
  \usepackage[dvips]{graphicx}
  \DeclareGraphicsExtensions{.eps}
\fi

\usepackage{amsmath}
\usepackage{upgreek}
\usepackage{array}
\usepackage{url}
\usepackage{siunitx}
\usepackage{textcomp,nicefrac}

	\usepackage{tikz}
	\usepackage{pgfgantt}

	\usetikzlibrary{decorations.pathreplacing}
	\usetikzlibrary{shapes,arrows,automata}
	\usetikzlibrary{positioning}
	\usetikzlibrary{decorations.markings}
	\usetikzlibrary{calc}
	\usetikzlibrary{fit}
	\usetikzlibrary{backgrounds}
	\usetikzlibrary{shapes,decorations,arrows,calc,arrows.meta,fit,positioning,shadows}
	\definecolor{redlight}{RGB}{235, 106, 132}
	\definecolor{greenlight}{RGB}{195, 252, 186}
	\definecolor{bluelight}{RGB}{178, 180, 249}

	\definecolor{goodbluebar}{RGB}{114, 147, 203}
	\definecolor{goodorangebar}{RGB}{225, 151, 76}
	\definecolor{goodgreenbar}{RGB}{132, 186, 91}
	\definecolor{goodredbar}{RGB}{211, 94, 96}
	\definecolor{goodblackbar}{RGB}{128, 133, 133}
	\definecolor{goodpurplebar}{RGB}{144, 103, 167}
	\definecolor{goodwinebar}{RGB}{171, 104, 87}
	\definecolor{goodgoldbar}{RGB}{204, 194, 16}

	\definecolor{goodblue}{RGB}{57, 106, 177}
	\definecolor{goodorange}{RGB}{218, 124, 48}
	\definecolor{goodgreen}{RGB}{62, 150, 81}
	\definecolor{goodred}{RGB}{204, 37, 41}
	\definecolor{goodblack}{RGB}{83, 81, 84}
	\definecolor{goodpurple}{RGB}{107, 76, 154}
	\definecolor{goodwine}{RGB}{146, 36, 40}
	\definecolor{goodgold}{RGB}{148, 139, 61}

	\definecolor{goodbluebar_origin}{RGB}{114, 147, 203}
    \definecolor{goodorangebar_origin}{RGB}{225, 151, 76}
    \definecolor{goodgreenbar_origin}{RGB}{132, 186, 91}
    \definecolor{goodredbar_origin}{RGB}{211, 94, 96}
    \definecolor{goodblackbar_origin}{RGB}{128, 133, 133}
    \definecolor{goodpurplebar_origin}{RGB}{144, 103, 167}
    \definecolor{goodwinebar_origin}{RGB}{171, 104, 87}
    \definecolor{goodgoldbar_origin}{RGB}{204, 194, 16}

	\tikzset{
	  photon/.style={decorate, decoration={snake}, draw=black},
	  fermion/.style={draw=black, postaction={decorate},decoration={markings,mark=at position .55 with {\arrow{>}}}},
	  vertex/.style={draw,shape=circle,fill=black,minimum size=3pt,inner sep=0pt},
	}
	\NewDocumentCommand\semiloop{O{black}mmmO{}O{above}}

	\newganttchartelement{orangebar}{
    orangebar/.style={
        inner sep=0pt,
        draw=black,
        very thick,
        top color=white,
        bottom color=goodorangebar
    },
    orangebar label font=\slshape,
    orangebar left shift=.1,
    orangebar right shift=-.1
	}

	\newganttchartelement{bluebar}{
		bluebar/.style={
			inner sep=0pt,
			draw=black,
			very thick,
			top color=white,
			bottom color=goodbluebar
		},
		bluebar label font=\slshape,
		bluebar left shift=.1,
		bluebar right shift=-.1
	}

	\newganttchartelement{greenbar}{
		greenbar/.style={
			inner sep=0pt,
			draw=black,
			very thick,
			top color=white,
			bottom color=goodgreenbar
		},
		greenbar label font=\slshape,
		greenbar left shift=.1,
		greenbar right shift=-.1
	}

    \tikzstyle{block} = [draw, fill=goodbluebar, rectangle, minimum height=1em, minimum width=10em]
    \tikzstyle{square} = [draw, fill=goodblackbar, rectangle]
    \tikzstyle{shadowsquare} = [draw, fill=goodblackbar, rectangle, double copy shadow={shadow xshift=-1ex, shadow yshift=1ex}]
    \tikzstyle{block_merger} = [draw, fill=goodredbar, rectangle, minimum height=1em, minimum width=10em]
    \tikzstyle{block_midas} = [draw, fill=goodgreenbar, rectangle, minimum height=1em, minimum width=10em]
    \tikzstyle{plate} = [draw, shape=rectangle, rounded corners=0.5ex, thick, dotted, inner xsep=15pt, inner ysep=20pt,label={[xshift=-4.75cm,yshift=6.75cm]south east:#1}]
    \tikzstyle{plate1} = [draw, shape=rectangle, rounded corners=0.5ex, thick, inner xsep=15pt, inner ysep=20pt, label={[xshift=-4.75cm,yshift=6.75cm]south east:#1}]

\def\BibTeX{{\rm B\kern-.05em{\sc i\kern-.025em b}\kern-.08em
T\kern-.1667em\lower.7ex\hbox{E}\kern-.125emX}}

\begin{document}

\title{Data Flow in the Mu3e DAQ}
\author{Marius K\"oppel$^{1}$ \IEEEauthorblockN{on behalf of the Mu3e collaboration}
\thanks{
This work has been submitted to the IEEE for possible publication.
Copyright may be transferred without notice, after which this version may no longer be accessible.
Corresponding author: M. Köppel (email: mkoeppel@uni-mainz.de).
}
\thanks{The author would like to thank the members of the Mu3e collaboration. The author is particularly grateful for the help of the different task-forces (Mu3e Integration Run task-force, Mu3e DAQ task-force, Mu3e Pixel task-force and MuTRiG task-force) from the Mu3e collaboration involved in this work. This work was supported by the PRISMA$^{+}$ (Precision Physics, Fundamental Interactions and Structure of Matter) Cluster of Excellence funded by the Deutsche Forschungsgemeinschaft (DFG, German Research Foundation) under Germany’s Excellence Strategy - EXC 2118 PRISMA$^{+}$ - 390831469.}
\thanks{Corresponding author: M. K\"oppel (email: mkoeppel@uni-mainz.de).}
\thanks{$^{1}$Institut f\"ur Kernphysik and PRISMA$^{+}$ Cluster of Excellence,
          Johannes Gutenberg-Universit\"at Mainz, Johann-Joachim-Becherweg 45, 55128 Mainz, Germany}
}

\maketitle

\begin{abstract}
The Mu3e experiment at the Paul Scherrer Institute (PSI) searches for the charged lepton flavour violating decay $\mu^+ \rightarrow e^+ e^+ e^-$.
The experiment aims for an ultimate sensitivity of one in $10^{16}$ $\mu$ decays. 
The first phase of the experiment, currently under construction, will reach a branching ratio sensitivity of $2\cdot10^{-15}$ by observing $10^{8}$ $\mu$ decays per second over a year of data taking.
The highly granular detector based on thin high-voltage monolithic active pixel sensors (HV-MAPS) and scintillating timing detectors will produce about~\SI{100}{Gbit/s} of data at these particle rates.
The Field Programmable Gate Array (FPGA) based Mu3e Data Acquisition System (DAQ) will read out the different detector parts.
The trigger-less readout system is used to sort, time align and analyse the data while running.
A farm of PCs equipped with powerful graphics processing units (GPUs) will perform the event reconstruction and data reduction.
This paper presents the ongoing integration of the sub detectors into the DAQ system, especially focusing on the time aligning and the data flow inside the FPGAs of the filter farm.
It discusses the DAQ system used in the Mu3e Integration Runs performed in spring 2021 and 2022, while providing results of the synchronisation of the different detectors.
\end{abstract}


\section{The Mu3e Experiment}

\IEEEPARstart{T}{he} Mu3e experiment \cite{ARNDT2021165679} at the Paul Scherrer Institute (PSI) searches for the charged lepton flavour violating decay $\mu^+ \rightarrow e^+ e^+ e^-$.
The experiment aims for an ultimate sensitivity of one in $10^{16}$ $\mu$ decays.
This would improve the current limit of $\mathcal B (\mu^+\rightarrow e^+e^+e^-) < \SI{1.0 e-12}{}$, set by the SINDRUM experiment~\cite{sindrum}, by four orders of magnitude.
The first phase of the experiment, currently under construction, will reach a branching ratio sensitivity of $2\cdot10^{-15}$ by observing $10^{8}\mu$ decays per second over a year of data taking.
The planned second phase depends upon a new High intensity Muon Beam-line (HiMB) \cite{himb} which will provide $10^{9}-10^{10}$ $\mu$ per second.

The~\SI{590}{MeV} proton beam at PSI hits a graphite target, producing secondary pions, which decay into muons.
The resulting~\SI{29.8}{MeV/c} muon beam is further guided into the Mu3e detector, sitting inside a~\SI{1}{T} solenoid magnet, and stopped by a hollow double cone target made out of Mylar foil.
These stopped muons decay at rest and trajectories of their decay products are measured using a silicon pixel tracking detector.

\begin{figure}[!t]
\centering
  \includegraphics[width=\linewidth]{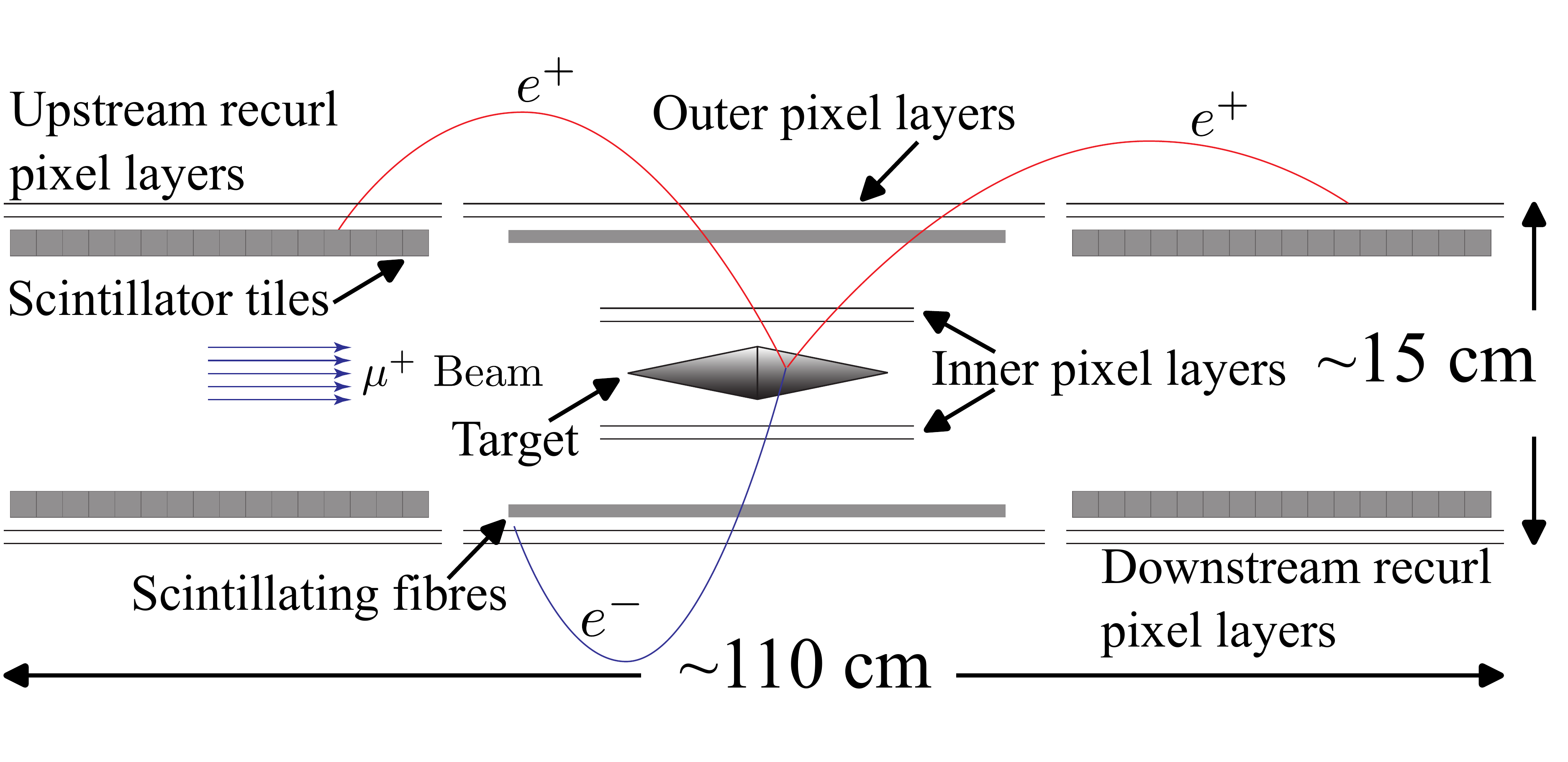}
  \caption{A sketch of the phase I Mu3e detector concept is shown. The \SI{110}{cm} long, cylindrical detector is cut along the beam axis. The incoming muons are stopped on a target and decay at rest. The target is surrounded by a two-layer vertex detector built out of MuPix pixel sensors. The second detector station holds scintillating fibres used for the timing measurement. Further out, a second double layer of pixel sensors is mounted. Additionally, two upstream and downstream recurl stations holding a third double layer of pixel sensors and scintillating tiles are used to improve the momentum and timing measurements for particles curling back in the \SI{1}{T} magnetic field.}
  \label{fig:detector}
\end{figure}
In Figure~\ref{fig:detector}, a sketch of the Mu3e detector is shown.
The detector is built of three parts, one inner central station and two recurl stations, one upstream and one downstream.
The recurl stations improve the momentum resolution by providing additional track points when the particles curl back towards the beam axis in the magnetic field.

Multiple scattering is the largest contribution to the momentum resolution.
Therefore, it is crucial to reduce the material budget.
Hence, a new type of thin high-voltage monolithic active pixel sensors (HV-MAPS), called MuPix, is used, which can be produced as thin as ~\SI{50}{\micro m}~\cite{mupix1,mupix2,mupix3,mupix4,mupix5,mupix6,mupix7}.
To have precise time-of-flight measurement and to suppress the background from random accidental coincidental decays, a precise timing measurement is needed.
Therefore, a scintillating fibre detector~\cite{fibre}, placed in the central station of the detector, and scintillating tile detectors~\cite{tile}, located in the two recurl stations, are used.
These detectors are read out via silicon photomultipliers and a custom-designed Application-Specific Integrated Circuit (ASIC), called MuTRiG~\cite{mutrig}.

To be able to handle the~\SI{100}{Gbit/s} of data, which are produced by the pixel and scintillating timing detectors, an online selection needs to be performed.
A farm of PCs equipped with powerful graphics processing units (GPUs) will perform the event reconstruction and data reduction.

\section{Mu3e Data Acquisition System}

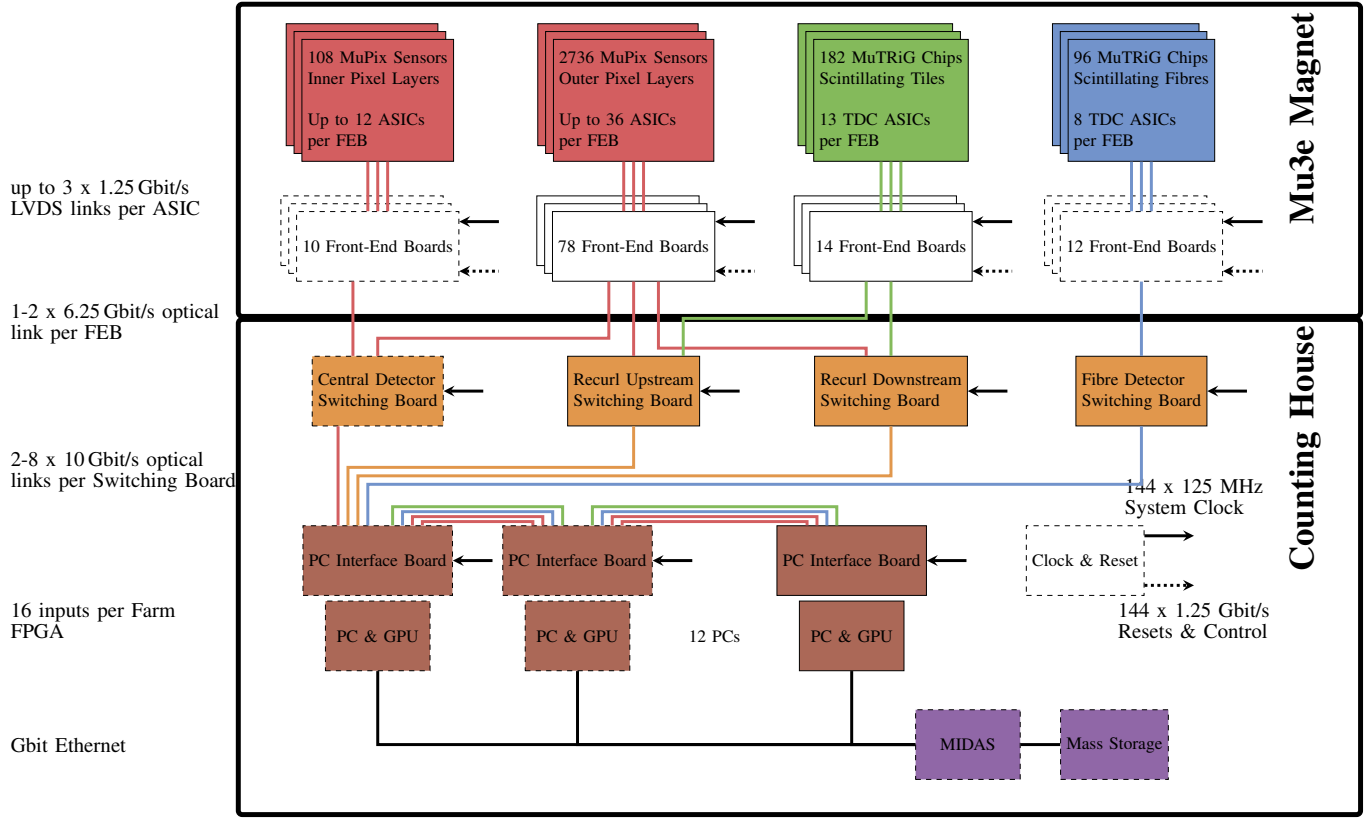
\begin{figure*}
    \centering
    \resizebox{\textwidth}{!}{%
    \begin{tikzpicture}[auto, node distance=2cm]

    \node [shadowsquare, minimum height=7em, minimum width=7em, fill=goodredbar, name=pix, align=left] (InnerPixel) {108 MuPix Sensors \\ Inner Pixel Layers  \\ \\ Up to 12 ASICs \\ per FEB};
    \node [shadowsquare, minimum height=7em, minimum width=7em, fill=goodredbar, name=pix2, right=2cm of InnerPixel, align=left] (Pixel) {2736 MuPix Sensors \\ Outer Pixel Layers \\  \\ Up to 36 ASICs \\ per FEB};
    \node [shadowsquare, minimum height=7em, minimum width=7em, fill=goodgreenbar, right=2cm of Pixel, name=til, align=left] (Tiles) {182 MuTRiG Chips \\ Scintillating Tiles \\ \\ 13 TDC ASICs \\ per FEB};
    \node [shadowsquare, minimum height=7em, minimum width=7em, fill=goodbluebar, right=2cm of Tiles, name=fib, align=left] (Fibre) {96 MuTRiG Chips \\ Scintillating Fibres \\ \\ 8 TDC ASICs \\ per FEB};

    \node [shadowsquare, minimum height=4em, minimum width=4em, below=1cm of InnerPixel, name=fpga1, fill=white, dashed, pattern color=goodblackbar] (FPGA1) {10 Front-End Boards};
    \draw[line width=0.6mm, stealth-] ([yshift=0.5 cm]FPGA1.east) -- +(0.8,0);
    \draw[line width=0.6mm, dotted, stealth-] ([yshift=-0.5 cm]FPGA1.east) -- +(0.8,0);
    \node [shadowsquare, minimum height=4em, minimum width=4em, below=1cm of Pixel, name=fpga1, fill=white] (FPGA2) {78 Front-End Boards};
    \draw[line width=0.6mm, stealth-] ([yshift=0.5 cm]FPGA2.east) -- +(0.8,0);
    \draw[line width=0.6mm, dotted, stealth-] ([yshift=-0.5 cm]FPGA2.east) -- +(0.8,0);
    \node [shadowsquare, minimum height=4em, minimum width=4em, below=1cm of Tiles, name=fpga2, fill=white] (FPGA3) {14 Front-End Boards};
    \draw[line width=0.6mm, stealth-] ([yshift=0.5 cm]FPGA3.east) -- +(0.8,0);
    \draw[line width=0.6mm, dotted, stealth-] ([yshift=-0.5 cm]FPGA3.east) -- +(0.8,0);
    \node [shadowsquare, minimum height=4em, minimum width=4em, below=1cm of Fibre, name=fpga3, fill=white, dashed] (FPGA4) {12 Front-End Boards};
    \draw[line width=0.6mm, stealth-] ([yshift=0.5 cm]FPGA4.east) -- +(0.8,0);
    \draw[line width=0.6mm, dotted, stealth-] ([yshift=-0.5 cm]FPGA4.east) -- +(0.8,0);

    \node (pointMagnet1) [left=0.5cm of InnerPixel] {};
    \node (pointMagnet2) [right=2cm of FPGA4] {};
    \begin{scope}[on background layer]
        \node[plate1=, fit=(pointMagnet1) (pointMagnet2) (InnerPixel) (FPGA4), line width=1mm] {};
    \end{scope}
    \node (text8) at (18.7,0) [text width=5cm, align=justify, rotate=90] {\huge \textbf{Mu3e Magnet}};

    \node [square, fill=goodorangebar, minimum height=4em, minimum width=5em, below=1.5cm of FPGA1, name=sw1, align=left, dashed] (SW1) {Central Detector \\ Switching Board};
    \draw[line width=0.6mm, stealth-] (SW1.east) -- +(0.8,0);
    \node [square, fill=goodorangebar, minimum height=4em, minimum width=5em, below=1.5cm of FPGA2, name=sw2, align=left] (SW2) {Recurl Upstream \\ Switching Board};
    \draw[line width=0.6mm, stealth-] (SW2.east) -- +(0.8,0);
    \node [square, fill=goodorangebar, minimum height=4em, minimum width=5em, below=1.5cm of FPGA3, name=sw3, align=left] (SW3) {Recurl Downstream \\ Switching Board};
    \draw[line width=0.6mm, stealth-] (SW3.east) -- +(0.8,0);
    \node [square, fill=goodorangebar, minimum height=4em, minimum width=5em, below=1.5cm of FPGA4, name=sw4, align=left] (SW4) {Fibre Detector \\ Switching Board};
    \draw[line width=0.6mm, stealth-] (SW4.east) -- +(0.8,0);


    \node [square, minimum height=4em, minimum width=5em, below=2cm of SW1, fill=goodwinebar, name=pcib1, dashed] (PCIB1) {PC Interface Board};
    \draw[line width=0.6mm, stealth-] (PCIB1.east) -- +(0.8,0);
    \node [square, minimum height=4em, minimum width=5em, right=1cm of PCIB1, fill=goodwinebar, name=pcib2, dashed] (PCIB2) {PC Interface Board};
    \draw[line width=0.6mm, stealth-] (PCIB2.east) -- +(0.8,0);
    \node [square, minimum height=4em, minimum width=5em, right=2.5cm of PCIB2, fill=goodwinebar, name=pcib3] (PCIB3) {PC Interface Board};
    \draw[line width=0.6mm, stealth-] (PCIB3.east) -- +(0.8,0);

    \node [square, minimum height=4em, minimum width=6em, below=0.1cm of PCIB1, fill=goodwinebar, name=pcib1, dashed] (PC1) {PC \& GPU};
    \node [square, minimum height=4em, minimum width=6em, below=0.1cm of PCIB2, fill=goodwinebar, name=pcib2, dashed] (PC2) {PC \& GPU};
    \node [square, minimum height=4em, minimum width=6em, below=0.1cm of PCIB3, fill=goodwinebar, name=pcib3] (PC12) {PC \& GPU};

    \node [square, fill=goodpurplebar, minimum height=4em, minimum width=6em, dashed] (server) at (11.9, -13) {MIDAS};
    \node [square, fill=goodpurplebar, minimum height=4em, minimum width=6em, right=0.8cm of server, dashed] (store) {Mass Storage};
    \node [square, minimum height=4em, minimum width=6em, right=2cm of PCIB3, fill=white, name=clock, dashed] (Clock) {Clock \& Reset};
    \draw[line width=0.6mm, -stealth, align=left] ([yshift=0.5 cm]Clock.east) -- +(1,0) node[above=0.2cm] {\large 144 x 125 MHz \\ \large System Clock};
    \draw[line width=0.6mm, dotted, -stealth, align=left] ([yshift=-0.5 cm]Clock.east) -- +(1,0) node[below=0.2cm] {\large 144 x 1.25 Gbit/s \\ \large Resets \& Control};

    \begin{scope}[->,>=latex]
        \foreach \i in {-1,0,1}{%

          \draw[line width=0.6mm,-, goodredbar] ([xshift=\i * 0.2 cm]InnerPixel.south) -- ([xshift=\i * 0.2 cm]FPGA1.north);

          \draw[line width=0.6mm,-, goodredbar] ([xshift=\i * 0.2 cm]Pixel.south) -- ([xshift=\i * 0.2 cm]FPGA2.north) ;}

        \foreach \i in {-1,0,1}{%
          \draw[line width=0.6mm,-, goodgreenbar] ([xshift=\i * 0.2 cm]Tiles.south) -- ([xshift=\i * 0.2 cm]FPGA3.north) ;}

        \foreach \i in {-1,0,1}{%
          \draw[line width=0.6mm,-, goodbluebar] ([xshift=\i * 0.2 cm]Fibre.south) -- ([xshift=\i * 0.2 cm]FPGA4.north) ;}

	\end{scope}

    \draw [line width=0.6mm, -, goodredbar] ([xshift=-0.8 cm]SW1.south) -- node {} ([xshift=-0.8 cm]PCIB1.north);
    \draw [line width=0.6mm, -, goodorangebar] (SW2.south) |- ($(SW2)!.45!(PCIB1)$) -| node {} ([xshift=-0.6 cm]PCIB1.north);
    \draw [line width=0.6mm, -, goodorangebar] (SW3.south) |- ($(SW2)!.5!(PCIB1)$) -| node {} ([xshift=-0.4 cm]PCIB1.north);
    \draw [line width=0.6mm,-, goodbluebar] (SW4.south) |- ($(SW2)!.55!(PCIB1)$) -| node {} ([xshift=-0.2 cm]PCIB1.north);

    \draw [line width=0.6mm,-, goodredbar] ([xshift=0.9 cm]PCIB1.north) |- (2,-8.5) -| node {} ([xshift=-0.9 cm]PCIB2.north);
    \draw [line width=0.6mm,-, goodredbar] ([xshift=0.7 cm]PCIB1.north) |- (2,-8.4) -| node {} ([xshift=-0.7 cm]PCIB2.north);
    \draw [line width=0.6mm,-, goodbluebar] ([xshift=0.5 cm]PCIB1.north) |- (2,-8.3) -| node {} ([xshift=-0.5 cm]PCIB2.north);
    \draw [line width=0.6mm,-, goodgreenbar] ([xshift=0.3 cm]PCIB1.north) |- (2,-8.2) -| node {} ([xshift=-0.3 cm]PCIB2.north);

    \draw [line width=0.6mm,-, goodredbar] ([xshift=0.9 cm]PCIB2.north) |- (7,-8.5) -| node {} ([xshift=-0.9 cm]PCIB3.north);
    \draw [line width=0.6mm,-, goodredbar] ([xshift=0.7 cm]PCIB2.north) |- (7,-8.4) -| node {} ([xshift=-0.7 cm]PCIB3.north);
    \draw [line width=0.6mm,-, goodbluebar] ([xshift=0.5 cm]PCIB2.north) |- (7,-8.3) -| node {} ([xshift=-0.5 cm]PCIB3.north);
    \draw [line width=0.6mm,-, goodgreenbar] ([xshift=0.3 cm]PCIB2.north) |- (7,-8.2) -| node {} ([xshift=-0.3 cm]PCIB3.north);

    \path (PC2) -- node[auto=false] {12 PCs} (PC12);

    \draw [line width=0.6mm,-, goodredbar] ([xshift=-0.5cm]FPGA1.south) -- node {} ([xshift=-0.5cm]SW1.north);

    \draw [line width=0.6mm,-, goodredbar] (FPGA2.south) -- node {} (SW2.north);
    \draw [line width=0.6mm,-, goodredbar] ([xshift=-0.5cm]FPGA2.south) |- (1,-4.8) -| node {} (SW1.north);
    \draw [line width=0.6mm,-, goodredbar] ([xshift=0.5cm] FPGA2.south) |- (8,-5) -| node {} ([xshift=-0.5cm] SW3.north);

    \draw [line width=0.6mm,-, goodgreenbar] ([xshift=-0.5cm]FPGA3.south) |- (8,-4.4) -| node {} ([xshift=1cm] SW2.north);
    \draw [line width=0.6mm,-, goodgreenbar] (FPGA3.south) -- node {} (SW3);

    \draw [line width=0.6mm,-, goodbluebar] (FPGA4.south) -- node {} (SW4);

    \draw [line width=0.6mm,-] (PC1) |- node {} (server);
    \draw [line width=0.6mm,-] (PC2) |- node {} (server);
    \draw [line width=0.6mm,-] (PC12) |- node {} (server);
    \draw [line width=0.6mm,-] (server) -- node {} (store);

    \node (text1) at (-4.9,-2)  [text width=5cm, align=justify] {\large up to 3 x~\SI{1.25}{Gbit/s} \\ LVDS links per ASIC};
    \node (text2) at (-4.9,-4.5)  [text width=5cm, align=justify] {\large 1-2 x~\SI{6.25}{Gbit/s} optical \\ link per FEB};
    \node (text3) at (-4.9,-7.5)  [text width=5cm, align=justify] {\large 2-8 x~\SI{10}{Gbit/s} optical \\ links per Switching Board};
    \node (text5) at (-4.9,-10.5)  [text width=5cm, align=justify] {\large 16 inputs per Farm \\ FPGA};
    \node (text5) at (-4.9,-13)  [text width=5cm, align=justify] {\large Gbit Ethernet};

    \node (pointCounting1) [below=5cm of pointMagnet1] {};
    \node (pointCounting2) [below=5cm of pointMagnet2] {};
     \begin{scope}[on background layer]
         \node[plate1=, line width=1mm, fit=(pointCounting1) (pointCounting2) (SW1) (store)] {};
     \end{scope}
    \node (text7) at (18.7,-7) [text width=5cm, align=justify, rotate=90] {\huge \textbf{Counting House}};

    \end{tikzpicture}
    }
    \caption{A sketch of the final Mu3e data acquisition system is shown. For the Integration Run 2021 and Cosmic Run 2022, the dashed parts of the system were used. The whole system is synchronised to a global \SI{125}{MHz} system clock using a clock and reset system holding the Genesys-2 FPGA board. This system is producing a \SI{1.25}{Gbit/s} reset and control line, which is used for run transitions. The inner vertex detector is read out via ten FEBs while the fibre detector is read out via two FEBs. The switching layer of the DAQ system holds one PCIe40 board (Switching Board) which is receiving the data from the FEBs, synchronises them and forwards it off to a minimal farm holding two Terasic DE5 boards (PC Interface Board).}
    \label{fig:intRunDAQ}
\end{figure*}

In Figure~\ref{fig:intRunDAQ}, the final Mu3e DAQ system is shown.
The total number of the MuPix sensors used for the phase I detector is 2844 while the scintillating fibres and scintillating tiles are read out by 278 MuTRiG ASICs.
The DAQ is designed to be able to handle the expected data rate of~\SI{100}{Gbit/s} for the full detector.
Both ASICs (MuPix and MuTRiG) send zero-suppressed and time unsorted hit data over up to three links for the MuPix and one link for the MuTRiG with a bandwidth of \SI{1.25}{Gbit/s} per link.

The MuPix chips from the inner pixel layer transmit up to \SI{30}{Mhits/s} over three links, while the outer layers are connected via one link.
The hit timestamp of the MuPix has a resolution of \SI{8}{ns}.

The MuTRiG chip generates a \SI{625}{MHz} coarse and fine counter with a bin size of \SI{50}{ps}, to ensure the high time resolution of less than \SI{500}{ps} for the fibre and less than \SI{100}{ps} for the tile detector \cite{ARNDT2021165679}.
Both counters are needed for the required timing resolution, but are generated in different ways on the MuTRiG chip.
In Section \ref{feb}, the special treatment of the coarse counter on the FEB is explained.

The actual DAQ system is built up of three layers of FPGA boards.
For reading out the different detectors, the custom developed Front-end boards (FEBs) are used.
This board is placed inside the magnetic field and is connected via~\SI{1.25}{Gbit/s} Low Voltage Differential Signaling (LVDS) links to the different sub-detectors.
The task of the board is to read out and configure the different detector ASICs, and sort the received hits in time.
The second layer of the DAQ system consists of the PCIe40 Board~\cite{pcie40} (in the paper the board will be called Switching Board), which was developed for the A Large Ion Collider Experiment (ALICE) and Large Hadron Collider beauty (LHCb) upgrades.
These boards are placed outside the magnet and are connected via~\SI{6.25}{Gbit/s} optical links to the FEBs.
They perform the time alignment of the different data streams, and they send the detector configuration via optical connections to the FEBs.
The last layer of the system consists of the commercial Terasic DE5a-Net-DDR4 boards~\cite{de5net} sitting in a PC equipped with a GPU (in the following, one of this units will be referred to as the farm PC).
The Terasic DE5a-Net-DDR4 boards hold an Arria10 FPGA and an onboard Double Data Rate Synchronous Dynamic Random-Access Memory (DDR SDRAM); in the following, this will be referred to as DDR4-RAM.
The DDR4-RAM is employed to buffer the data of the full detector, while the hits of the four layers of the central pixel detector are used to perform an online event selection on the GPU.
The total filter farm of the final system contains twelve units of farm PCs, which are daisy-chained to process the expected data rate of~\SI{100}{Gbit/s} from the detector.
To synchronise the different detectors, a dedicated clock and reset system was built.
The system is centralised around the Genesys-2 FPGA board~\cite{Genesys}, which generates a~\SI{125}{MHz} clock and synchronised reset signals for run starts and stops.
Additional electronics are used to create 144 copies of the clock and reset signals, which are connected to each part of the DAQ system.
A comprehensive description of the final DAQ of the Mu3e detector is given in~\cite{daq}.

Since the search for $\mu^+ \rightarrow e^+ e^+ e^-$ is a three body decay at rest, the DAQ system needs to read out the full detector in order to be able to select physics events.
This is in contrast to trigger-based DAQ systems, which only read out data if a certain detector response is over a specific threshold.
Examples of trigger based DAQ systems are ATLAS~\cite{atlasdaq} and Compact Muon Solenoid Experiment (CMS)~\cite{cmsdaq}, the two largest experiments at the Large Hadron Collider.
However, the concept of full online reconstruction of the whole detector data is also being implemented at the LHCb~\cite{lhcbDaq} upgrade, and for parts of the ALICE detectors~\cite{alice}.

The main contributions of this work, in contrast to the previous description of the Mu3e DAQ system~\cite{daq}, are:
\begin{enumerate}
  \item Detailed insights in the developed firmware parts of the Mu3e DAQ system, which are used to process the data from the different detectors to the filter farm.
  \item Description of the time alignment firmware used on the switching board (SWB) to synchronise the different detector systems.
  \item Overview of the Mu3e Cosmic Run 2022 and first timing studies of the different detectors using a vertical slice of the Mu3e DAQ system.
\end{enumerate}

\section{Data Flow In The Mu3e DAQ}

\subsection{Front-end Board}\label{feb}

\begin{figure}
  \centering
  \includegraphics[width=\linewidth]{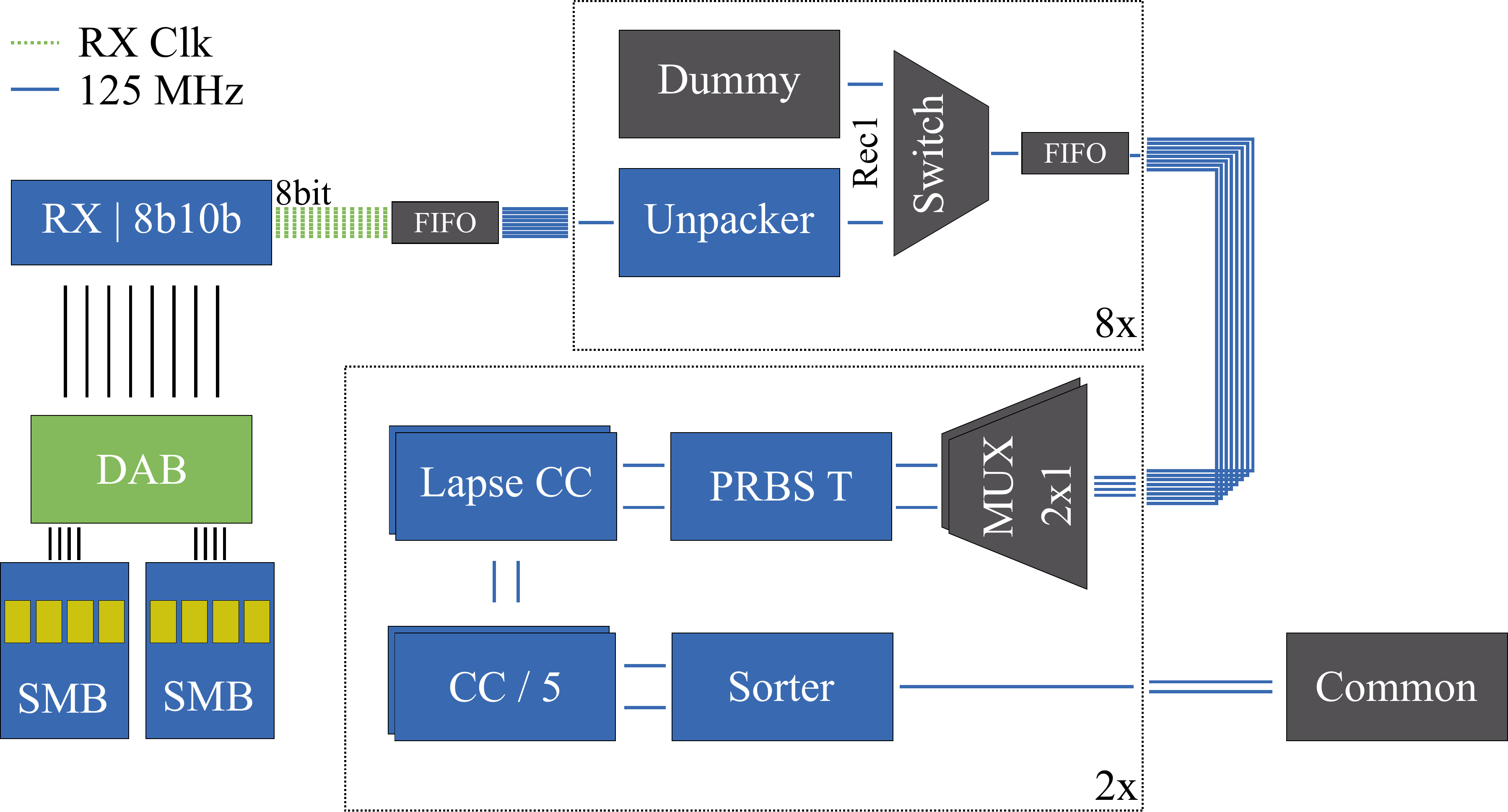}
  \caption{Sketch of the datapath of the scintillating fibre detector. One scintillating fibre (SciFi) module board (SMB) holds four MuTRiG ASICs, and is connected to the FEB via a detector adapter board (DAB). The receiver (RX) on the FEB is byte-aligning the incoming serial LVDS links and 8b/10b~\cite{8b10b} decoding them. Afterwards, the data is buffered in First In First Out (FIFO) queues and unpacked, multiplexed, and time-corrected. After this pre-processing, the hit data can be sorted using onboard memory, and transmitted to the common part of the FEB.}
  \label{fig:feb}
\end{figure}
As an example for the developed data processing firmware on the FEB, the data path of the scintillating fibre detector is shown in Figure~\ref{fig:feb}.
Two scintillating fibre (SciFi) module boards (SMBs) holding eight MuTRiG ASICs are connected, via a specific detector adapter board (DAB), to a FEB.
After receiving the data on the Arria V FPGA, the hit data is unpacked and grouped into a detector-specific record type (Rec1).
The data lines from the different ASICs are merged into groups of two for further processing.

To be able to correlate the hits from the different detectors, time sorting is needed. Unfortunately, the two ASICs create the hit time with two different frequencies.
The timestamp generation of the MuTRiG chip runs at \SI{625}{MHz} while the MuPix chip generates the timestamp with \SI{125}{MHz}.
The faster clock for the MuTRiG is needed for the required time resolution of the two timing detectors.
To encounter this problem the firmware needs to provide a common time base for both detectors.
Therefore, the firmware takes the timestamp from the MuTRiG hits and divides them into a \SI{125}{MHz} and a \SI{625}{MHz} part.
Afterwards the firmware sorts the hits according to the \SI{125}{MHz} part.
Since the MuPix chips are running already at \SI{125}{MHz} only the sorting is needed.

In detail the algorithm for sorting the MuTRiG hits into \SI{125}{MHz} is a bit more involved.
In order to run at \SI{625}{MHz}, the coarse counter timestamp is implemented as an optimised fifteen-stage linear feedback shift register, which runs through a deterministic order of $2^{15}-1$ stages.
The feedback loop of the shift register is implemented using an exclusive-or (XOR).
These stages can appear to be pseudo random values (e.g. 1. stage is 0x1234, 2. stage 0x5678, 3. stage 0x4242), but can be translated back to binary values (e.g. 0x1234 = 0, 0x5678 = 1, 0x4242 = 2) using a dual-port lookup RAM on the FEB (in Figure~\ref{fig:feb}, this is indicated with the PRBS T block).
Dual-port RAMs are used to be able to process the two data streams at the same time.

Since the timestamp of the MuPix chip counts up to $2^{15}$ and the coarse counter of the MuTRiG chip counts up to $2^{15}-1$, the counting of the MuTRiG chip needs to be corrected.
As shown in \ref{fig:intRunDAQ} the whole system (FPGA-Boards and ASICs) is running in sync to a global \SI{125}{MHz} clock.
Both the \SI{125}{MHz} clock on the FEB and the \SI{625}{MHz} clock on the MuTRiG are synchronised to the global \SI{125}{MHz} clock.
Because of this synchronisation the lapse correction block (Lapse CC) can mimic the counting of the MuTRiG ASIC.
It can then estimate the number of counter overflows and subtracted them from the coarse counter to correct it for counting up to $2^{15}$.

After this correction of the counter value, the \SI{625}{MHz} counter needs to be divided by five to have a part which follows the \SI{125}{MHz} clock and a remainder part which holds the \SI{625}{MHz} value.
After all the corrections, the various hits can be sorted and transmitted to the SWB via up to two \SI{6.25}{Gbit/s} optical links (for the pixel and tile detectors only one optical link is needed).

\subsection{Switching Board}

\begin{figure}
  \centering
  \includegraphics[width=\linewidth]{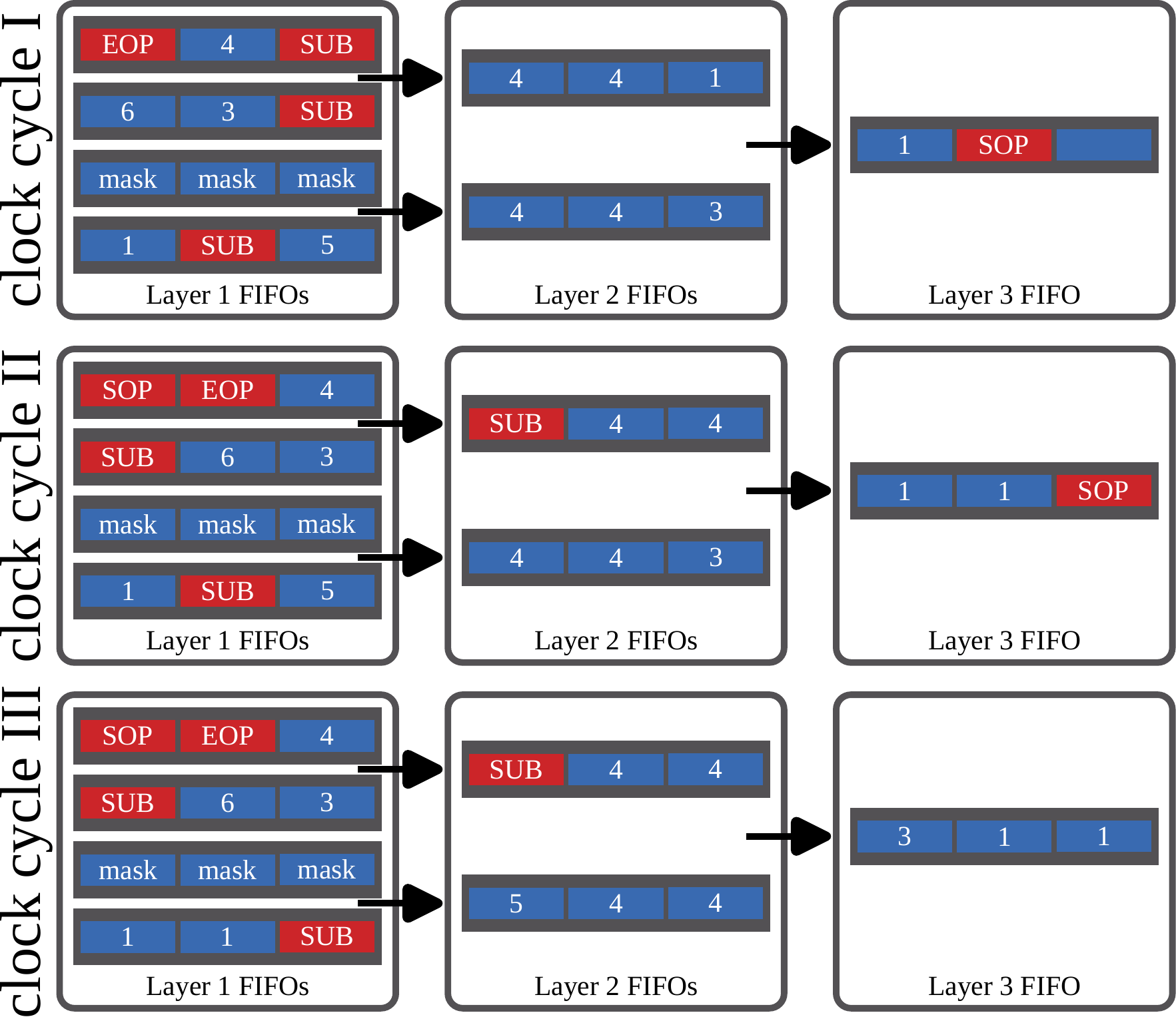}
  \caption{Example of a four-to-one time alignment firmware on the SWB using pairwise comparison in a tree. The numbers on the left indicate three example steps (clock cycles) of the algorithm. The markers SOP and EOP indicate the start and the end of a data package sent from the FEBs. The marker SUB shows the so-called sub header which contains more timestamp information. The values between show the hit time of hits inside a sub header. The alignment firmware combines the various packages into one common stream while keeping the hits' time sorted.}
  \label{fig:swb}
\end{figure}
One task of the SWB is to time-align the different data streams.
The time alignment of the hit data is needed to be able to perform the full track reconstruction on the GPU.
In Figure~\ref{fig:swb}, a sketch of this time alignment firmware of the SWB is shown.
Each data package from the different FEBs contains hit data sorted into \SI{8}{ns} bins.
A new data package is sent every \SI{16}{\micro s}.
Each hit contains space, time-over-threshold (ToT) of the comparator, and time information.
The data package for each detector is designed to hold the whole hit information using \SI{32}{bit}.
To be able to keep the size of \SI{32}{bits}, we only have four bits for the time information (in Figure~\ref{fig:swb} the time information of the \SI{32}{bit} hit word is the white number of the blue rectangles).
Therefore, we have to send 128 sub headers (SUB) which contain the upper seven bits of the global timestamp (in total one package as $2^{11}$ possible timestamps) when the four bits of the hit time overflow.

The time alignment firmware is using pairwise comparison of the hit timestamps in a tree-based architecture.
In Figure~\ref{fig:swb}, the start of each package is marked with SOP and the end of each package is marked with EOP.
Each layer of the tree contains two input streams and one output stream.
The streams are buffered in FIFO queues, the example in Figure~\ref{fig:swb} uses a queue size of three (in the final system the queue size is of the order of $2^{10}$).
In the first layer, the input stream is running with \SI{125}{MHz} and the output of the FIFO is running at \SI{250}{MHz}.
At each layer of the tree, both streams are buffered until each has a SOP, then the hit with the lowest timestamp is sent to the next layer.
If a stream holds a SUB, the hits from the other stream are forwarded until both streams hold the same SUB. Since the FEB always sends all 128 SUBs for each package - even in the case where there are no hits for these timestamps - the firmware only needs to compare the four bits of the two individual hits.
These are reproduced over the full tree until the EOP marker is reached.
Figure~\ref{fig:swb} shows, as an example, three steps (clock cycles) of a four-to-one tree.
The actual implementation is done with an eight-to-one tree following the same principle.
After the synchronisation, the merged stream is sent via a \SI{10}{Gbit/s} link to the first Terasic DE5 board of the filter farm.

As described above, the firmware combines eight input streams into one output stream.
Since there is only one clock frequency change from \SI{125}{MHz} to \SI{250}{MHz}, the time alignment firmware has a bottleneck of effectively $4-1$.
Using simulated data, the average rate for the whole pixel detector was estimated to be \SI{56}{Gbit/s}, including \SI{8}{bit}/\SI{10}{bit} encoding and \SI{75}{\%} protocol efficiency \cite{daq}.
The total number of input links for the central detector SWB from the FEBs is 36, while there is a total of eight output links to the Terasic DE5 board.
Therefore, the possible throughput of the time alignment firmware for the central detector SWB is of the order of~\SI{64}{Gbit/s}.
This reduces the bottleneck of the time alignment firmware to around $2-1$ for the central pixel detector.
However, this bottleneck is only hypothetical and only if all 36 input links of the central pixel SWB are 100\% saturated which will never be the case in the actual detector.
The expected rate of the whole pixel detector is \SI{56}{Gbit/s}, including the two recurl stations and the central station.
Since the central SWB is only connected to the central pixel detector, the time alignment firmware does not have to process the full \SI{56}{Gbit/s}.
To be able to use backpressure and to handle possible rate bursts, the FIFO queue size for the first layer of the time alignment firmware was set to hold one data package (\SI{32}{kB}).
For the two recurl SWBs, two \SI{10}{Gbit/s} links to the Terasic DE5 board is intended.
The recurl pixel stations produce only a fraction of the data from the central station, leading to no significant bottleneck.
For the other two detectors, the overall situation is relaxed as well.
The fibre detector will produce an overall rate of \SI{28}{Gbit/s} and one SWB is processing this data.
For the connection from the fibre detector SWB to the Terasic DE5 board four \SI{10}{Gbit/s} links are provided.
The two stations of the tile detector producing an overall rate of \SI{17}{Gbit/s}, while each recurl SWB has a \SI{10}{Gbit/s} link for the tile detector to the Terasic DE5 board.

\subsection{Filter Farm}

\begin{figure}
  \centering
  \includegraphics[width=\linewidth]{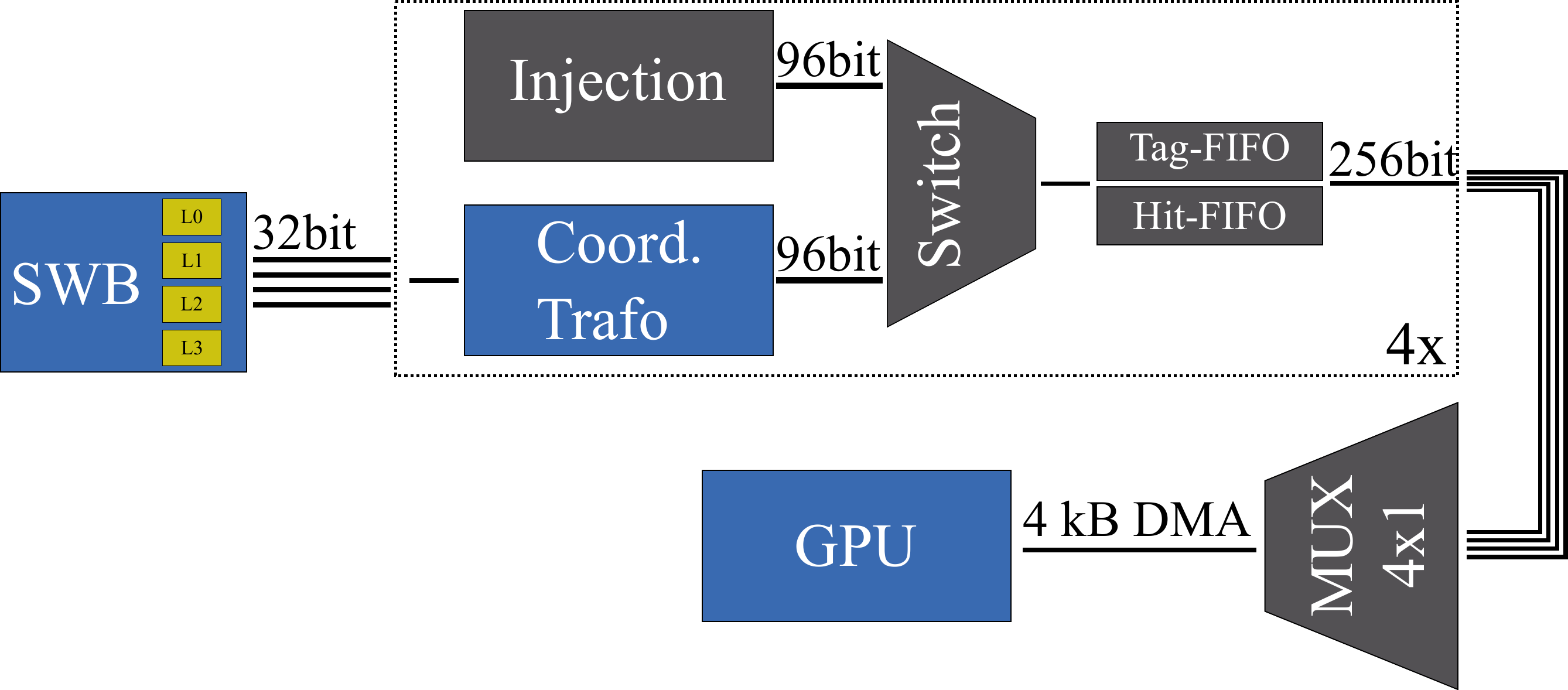}
  \caption{The central switching board (SWB) is connected to the first farm PC, sending the pixel hits of the four layers (L0-L3). The incoming hit positions are converted to \SI{32}{bit} floating-point values of the global $x,y,z$ position. For debugging and eventual blinding proceedings, an injection entity is implemented. Using a Tag-FIFO the absolute number of hits and total number of \SI{256}{bit} words for all hits inside a time-frame is stored. By multiplexing between the four layers, the required GPU events are contracted and sent via DMA to CPU RAM, where it is forwarded to the GPU.}
  \label{fig:gpu}
\end{figure}

The first Terasic DE5 board of the filter farm receives the whole \SI{100}{Gbit/s} detector data from all four SWB.
Because all data is synchronised and sorted to \SI{8}{ns}, the board receives a stream of time-continuous data.
This data is buffered on one of the two onboard memory interfaces.
If the buffer gets full, the data is forwarded to the next Terasic DE5 board which is daisy-chained.
Since the full farm contains twelve farm PCs, each of them needs to process only $\nicefrac{1}{12}$ of the overall \SI{100}{Gbit/s} rate.
To perform the online track reconstruction, the data of the central pixel detector needs to be transferred to the GPU which is hosted in the same PC.
The central pixel data is packed into \SI{2}{MB} packages and transmitted via the PC RAM to the GPU using direct memory access (DMA)~\cite{bruch,valentin}.
A reference of the selected events are sent back to the Terasic DE5 board using Peripheral Component Interconnect Express (PCIe) registers and stored in a request FIFO to trigger the readout of the data buffer.
By using two DDR4 RAMs, one can be used to buffered the data while the other one can be read out via a second DMA engine and transferred to the Maximum Integration Data Acquisition System (MIDAS)~\cite{midas1,midas2} which is used as data acquisition software.

To be able to efficiently perform the tracking on the GPU, the data needs to be prepared on the FPGA.
In Figure~\ref{fig:gpu}, the data flow of the central pixel detector on the Terasic DE5 board is shown.
Each layer sends its individual hit position and timing information using a total of \SI{32}{bits}.
In the first step, the position is converted into global \SI{32}{bit} floating point values of $x,y,z$ using \SI{96}{bits} in total.
For this, a conversion from the local coordinate system on the chip to the global coordinate system of the detector needs to be performed.
This is done by using the chip ID as address to lookup the global corner value (row=0, column=0) of the corresponding pixel chip in a RAM-based lookup-table.
Knowing the corner value of each chip and the column (col) and row information the global position $x,y,z$ in the whole pixel detector of the $i$th chip can be calculated using:

\begin{equation}
     \vec{h_i} = \vec{s_i} + \text{col}_i \cdot \vec{c_i} + \text{row}_i \cdot \vec{r_i},
\end{equation}
where $\vec{h_i}$ is the hit position vector $(x,y,z)$, $\vec{s_i}$ is the global corner position, $\vec{c_i}$ and $\vec{r_i}$ are the column and row directions.
Following that, the hit data is packaged into frames of 8ns and packed into \SI{256}{bits} to be processed by the DMA engine running at \SI{256}{bits} times \SI{250}{MHz}.
For each GPU frame, the hits are sorted in time for each of the four layers.
The hits of the four layers are stored in \SI{0.5}{MB} sub-packages per layer inside the overall \SI{2}{MB} package.
At the end of each sub-package, references to the \SI{8}{ns} borders of the hits are transmitted.
With this packaging, the GPU is able to dynamically create time frames in multiple of \SI{8}{ns}.
Since one has the reference on the \SI{8}{ns} level, it is possible to include hits from neighbouring frames ($\pm$ n x \SI{8}{ns}) without copying them twice.
These overlaps are needed to perform the tracking using all four layers~\cite{bruch,valentin}.

\section{The Mu3e Cosmic Run 2022}
\label{sec:intRun}

\begin{figure}
  \centering
  \includegraphics[width=\linewidth]{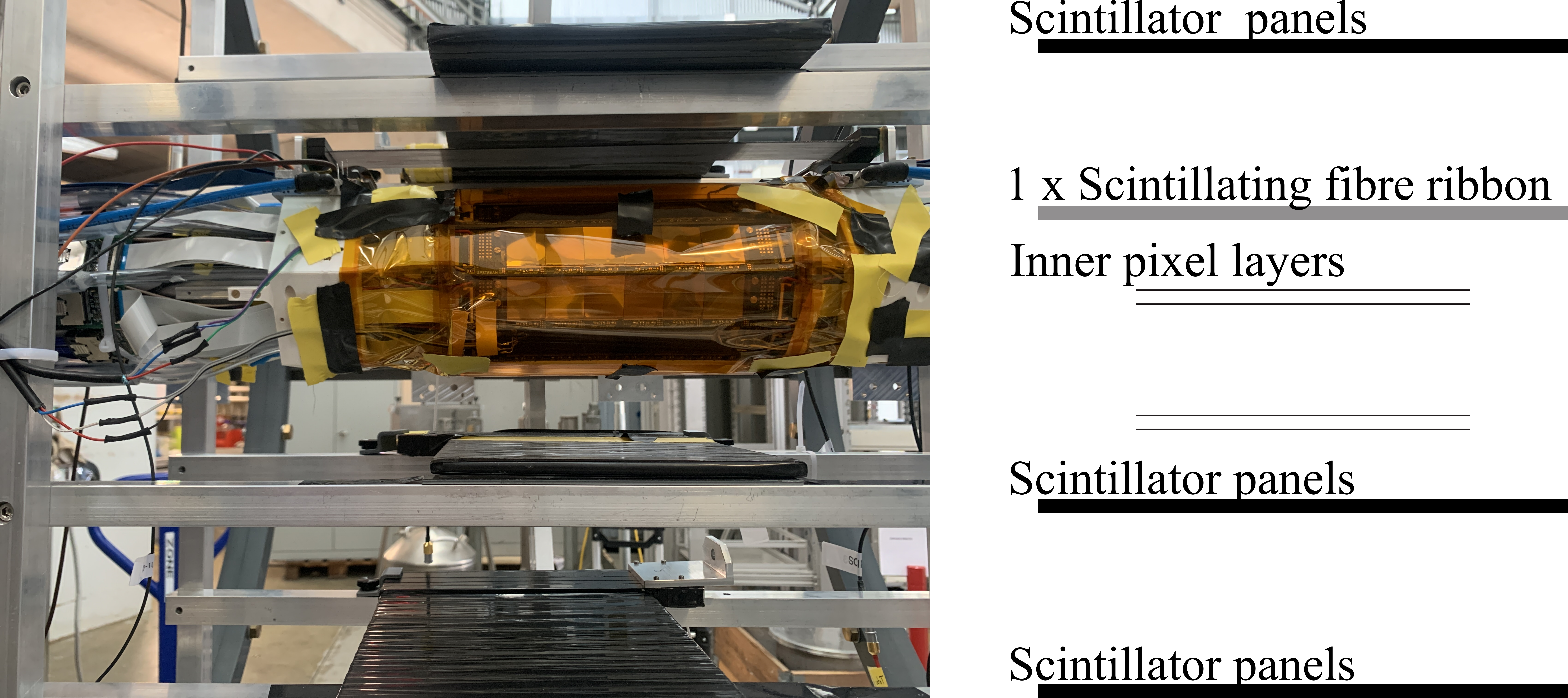}
  \caption{\textbf{Left:} Picture of the Mu3e Cosmic Run 2022 detector prototypes. The inner vertex detector is enclosed with Kapton foil to distribute the helium flow. Directly over the pixel detector, a layer of scintillating fibre ribbons is mounted. For further noise and cosmic studies, three additional scintillating panels are mounted around the detectors. The coincidence of the panels are used as a cosmic ray trigger. The three panels are not perfectly aligned, which results in a non optimal coverage of the detector acceptance.
  \textbf{Right:} Sketch of the detector setup of the Mu3e Cosmic Run 2022.}
  \label{fig:intRun}
\end{figure}

\begin{figure}
  \centering
  \includegraphics[width=\linewidth]{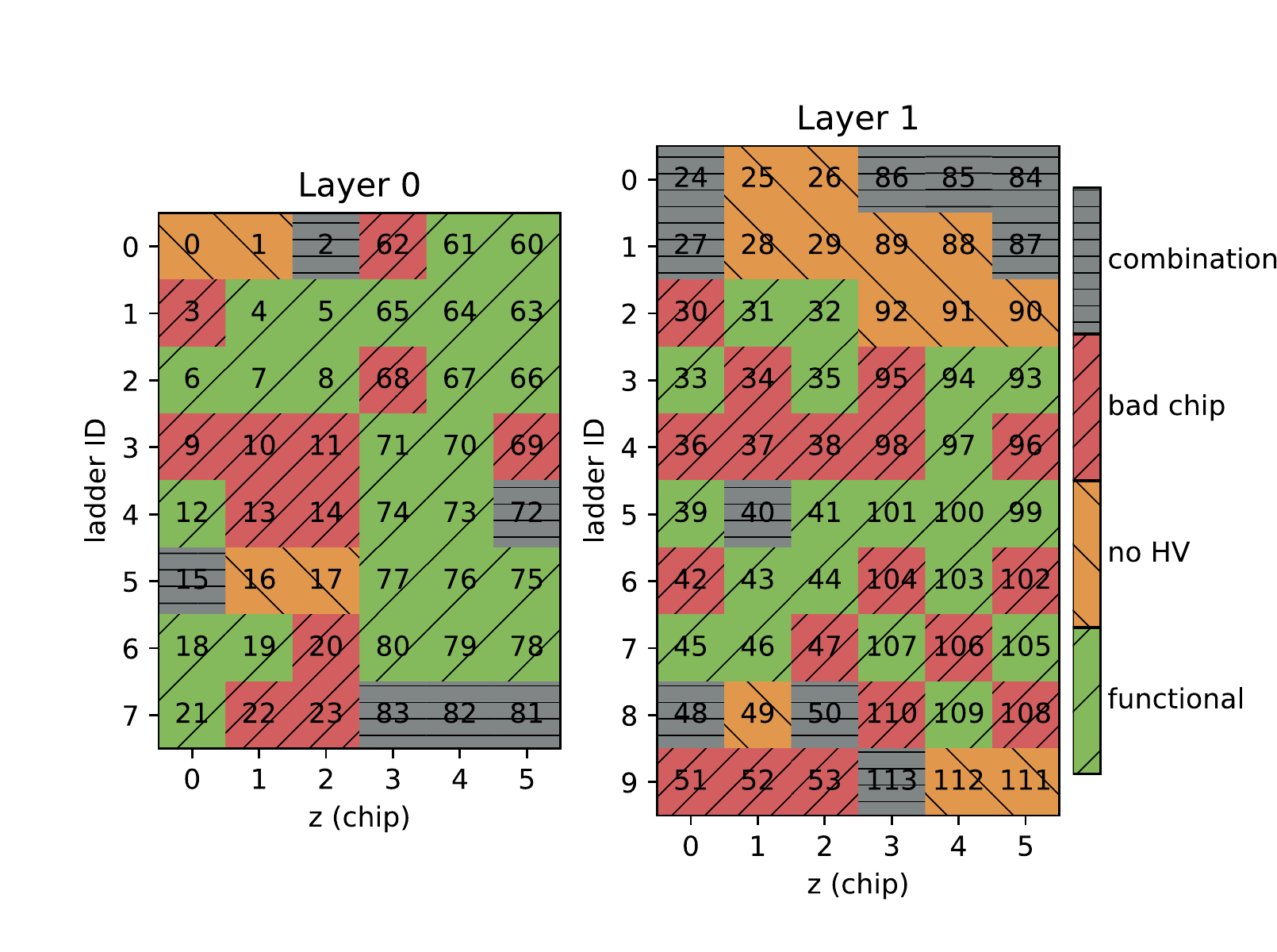}
  \caption{Overview of the chips mounted on the two layers of the inner vertex detector. All chips without problems are marked in green (with a wide bottom left to top right hatching), while the chips in orange (with a wide top left to bottom right hatching) run into thermal runaway when applying high voltage. The chips marked in red (with a fine bottom left to top right hatching) had problems in terms of broken links, non-working LVDS connections or noisy pixels. The grey chips (with a fine left to right hatching) had some combination of the two problems above.}
  \label{fig:workingchips}
\end{figure}

\begin{figure*}
  \centering
  \includegraphics[width=\linewidth]{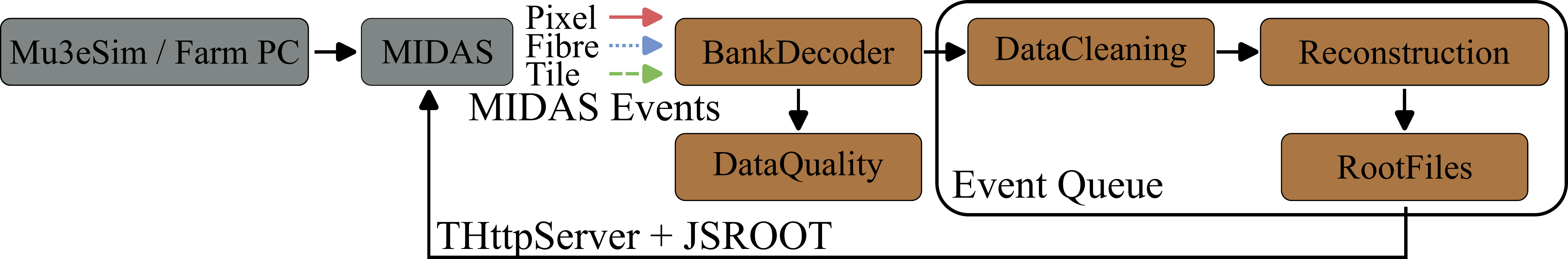}
  \caption{Sketch of the online analyzer framework used at the Mu3e Cosmic Run 2022. The design of the system is completely integrated into MIDAS and receives all the different detector events. In the first step, the data banks inside the events are decoded and sent to a queue where different cleaning and clustering steps are performed for each detector type. Using the Mu3e Reconstruction software, online tracking can be performed. The final tracks are written to a ROOT~\cite{root} file. At the same time, the raw data is buffered using a second queue to perform data quality plots (DataQuality) of the system. Using the THttpServer class of ROOT combined with JavaScript ROOT, provides interactive ROOT-like graphics in the web browsers. For debugging and performance tests, the whole setup can be fed with simulation data from the Mu3e software~\cite{trirec}.}
  \label{fig:analyzer}
\end{figure*}

\begin{figure}
\centering
\includegraphics[width=0.4\textwidth]{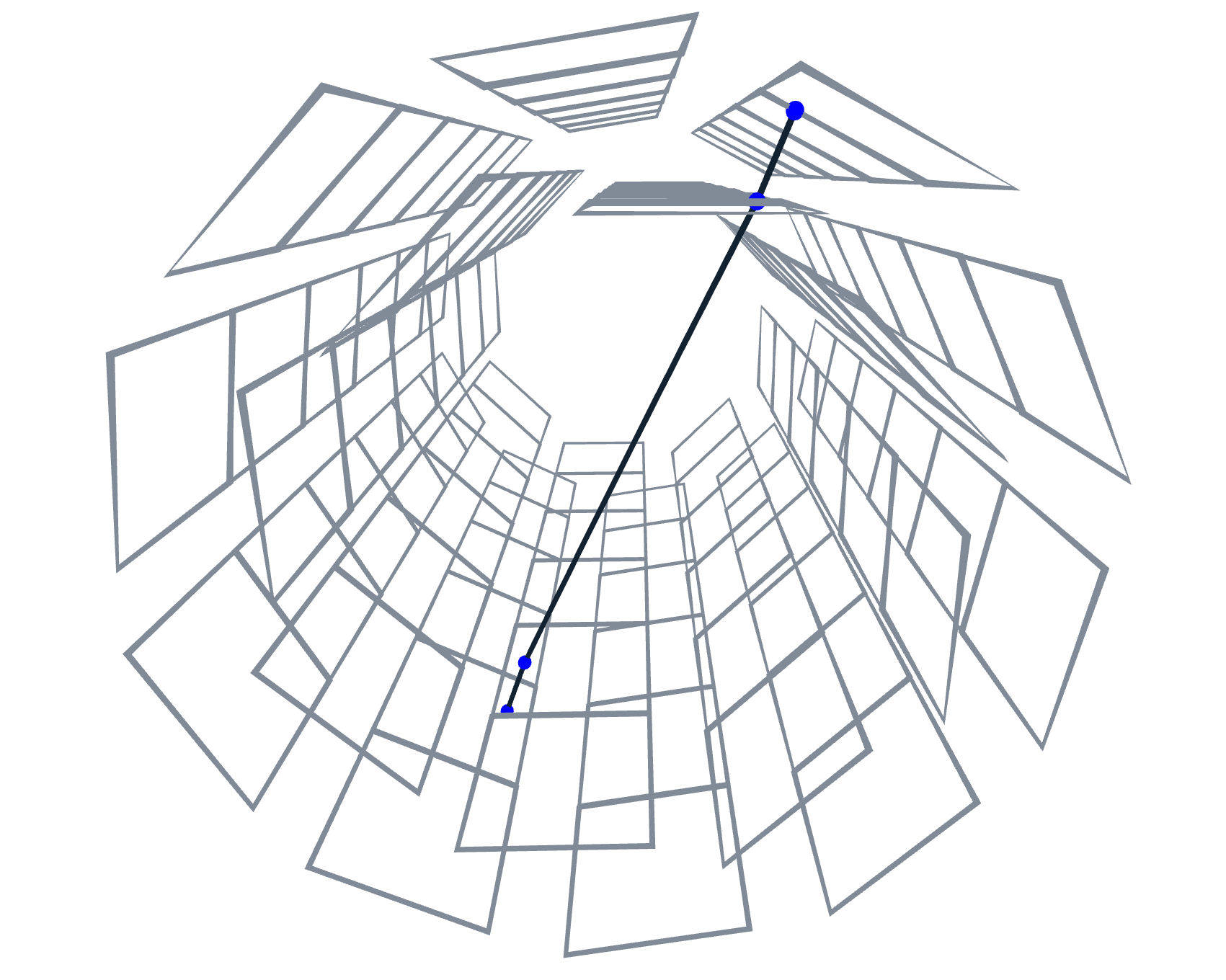}
\caption{Sketch of the Mu3e Event Display (PhD thesis in progress: B. Gayther, Preparations for Phase I of the Mu3e experiment). The display is showing a track which was reconstructed using the Mu3e Online Analyzer. The track was fitted to four hits in a certain time frame using only the data from the inner vertex detector.}
\label{fig:event_display}
\end{figure}

During the integration of the different detector systems into the final DAQ system, two intensive system tests were performed.
The first one, the Mu3e Integration Run 2021, was performed to integrate the inner vertex detector into the DAQ system and is explained in detail in \cite{intrun2021}.
The second test run, the Mu3e Cosmic Run 2022, was performed to finalise the integration of the inner vertex detector and to integrate the scintillating fibre detector using cosmic rays.
In contrast to the Integration Run 2021, the detectors were not operated inside a magnetic field.
In Figure~\ref{fig:intRun}, the Cosmic Run setup is shown.
One intention of the run was to operate the detectors under conditions as close as possible to the final experiment and study their noise behaviour.
This includes cooling of the MuPix detector with gaseous helium.

As for the final DAQ system, the Mu3e Cosmic Run 2022 contained ten FEBs for reading out the inner vertex detector.
On the other side, only two FEBs for the fibre detector were used compared to the final twelve.
Furthermore, there were no tiles and none of the other pixel detectors present.
All twelve FEBs are read out with only one SWB, which contains a third of the final links for the central vertex detector and a small fraction for the fibre detector.
The farm contained only two Terasic DE5 boards and no GPUs, in contrast to the twelve Terasic DE5 boards used in the final system.
The dashed blocks in Figure \ref{fig:intRunDAQ} show the components used in the Mu3e Cosmic Run 2022.
Furthermore, three layers of scintillator panels are mounted around the scintillating fibre and inner vertex detectors.
Using Nuclear Instrumentation Module (NIM) logic, the three panels are used to create a cosmic ray coincidence trigger.
This trigger signal was digitised via an additional FEB and sent to the SWB.
The geometry of the scintillator panels was not optimal, and the panels did not cover the whole detector acceptance.

By tuning the vertex detector to a minimal noise level, it was possible to read out the full detector data.
Due to a broken readout board, only one side of the fibre detector was operating.
In addition, not all pixel chips were able to run under optimal conditions.
In Figure \ref{fig:workingchips}, all chips without problems are marked in green, while the chips in orange run into thermal runaway when applying high voltage.
The chips marked in red had problems in terms of broken links, non-working LVDS connections or noisy pixels.
The grey chips had some combination of the two problems above.
Overall, only \SI{10}{V} could be applied to the chips since there are unforeseen problems in the prototype of the MuPix chip which are resolved in a newer version.
Having all these problems even the best-working chips were not operated at design conditions, leading to a worse timing resolution.
A detailed description of the inner vertex detector used in the test runs can be found in~\cite{thomas}.

\subsection{Online Analyzer}

The Mus3 Online Analyzer software was implemented to analyse the data quality and perform tracking.
This software component is based on the MIDAS-compatible manalyzer framework~\cite{manalyzer}.
A parallel readout pipeline was used to clean the data and perform the needed quality checks.
Figure~\ref{fig:analyzer} demonstrates the various steps of the analyzer.
In the first part, the system connects to MIDAS and checks if detector events are present.
The different detector events for the fibre and the vertex detector are decoded and sent to a queue for further processing.
Even with the threshold tuning and masking pixels, a few chips still had noisy pixels left which could not be turned off in hardware.
This required to filter them out of the raw data in software.
Since the fibre data path had no hardware sorter running, the data were sorted in time to be able to correlate with the pixel data.
The additional scintillator panels were considered as pixel hits and were filtered out at the pixel readout chain.
After the cleaning, a full track reconstruction was performed using an adapted version of the Mu3e triplet reconstruction software~\cite{trirec} to reconstruct cosmic tracks in the pixel detector.
The reconstructed hits were forwarded to an event display to have a live view on specific events (see Figure~\ref{fig:event_display}).
For general time correlations, the scintillator panels were used as a trigger to correlate the fibre and pixel hits in a time-window around the trigger.
Simple quality plots such as hit rate, hitmaps, timestamp plots, etc., were created for the different detectors.

\subsection{Results Of The Cosmic Run 2022}

\begin{figure}[t]
\includegraphics[width=0.49\textwidth]{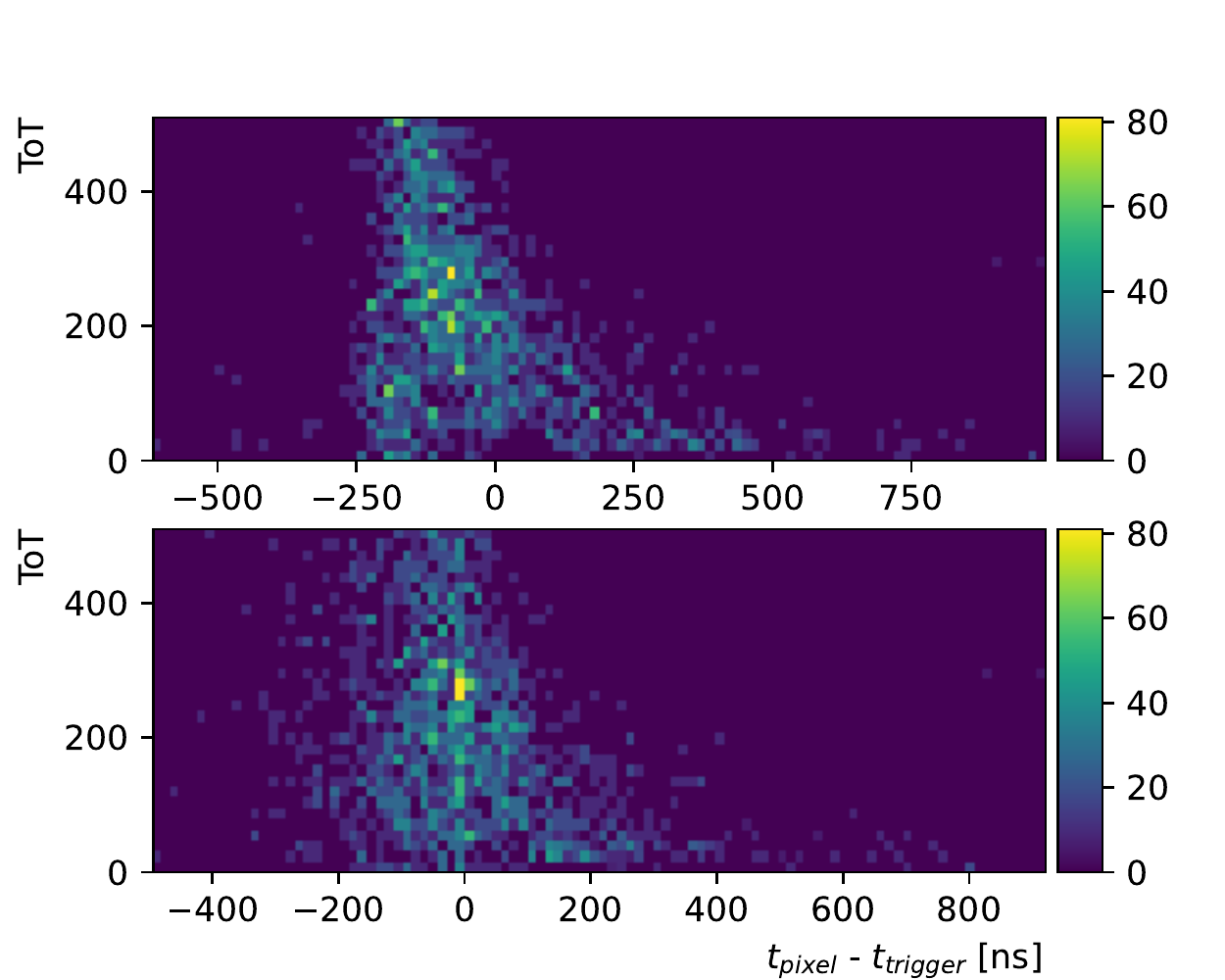}
\caption{\textbf{Top:} ToT versus time difference of the inner vertex detector and the scintillating panels. \textbf{Bottom:} ToT versus time difference of the inner vertex detector and the scintillating panels with time walk corrected pixel hit time.}
\label{fig:timewalk}
\end{figure}

\begin{figure}
\includegraphics[width=0.49\textwidth]{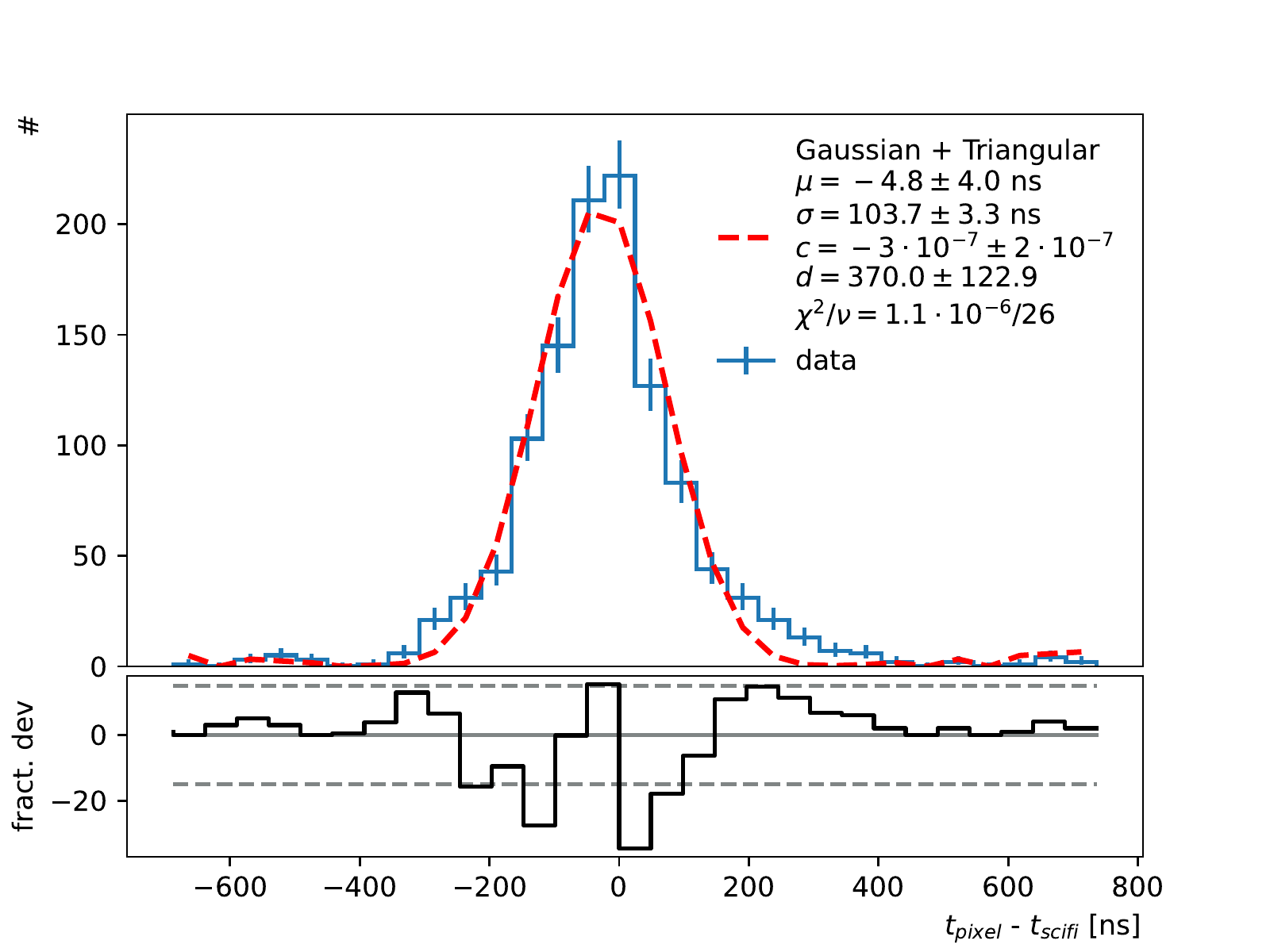}
\caption{The time correlation between the inner vertex detector and the scintillating fibre detector requires a coincidence of the scintillator panels.}
\label{fig:coincidences}
\end{figure}

To simplify analyses of the time correlation of the inner vertex detector and the scintillating fibre detector, only the hits which met a coincidence with the three scintillator panels are used.
The top plot of Figure~\ref{fig:timewalk} shows the ToT versus the time difference of the inner vertex detector and the scintillator panels.
The clear asymmetric shape of the distribution is caused by the time walk effect.
Since the comparator of each pixel cell has an absolute threshold on the rising edge of the analog signal to determine the hit time, signals with higher amplitude have an earlier time, and signals with lower amplitude have a later time.
The difference of the two times is referred to as time walk.
To overcome this problem, the MuPix chips can be tuned to use a second timestamp, sampled on the falling edge of the signal, to calculate the ToT and correct the hit time.
During the Cosmic Run this feature was not turned on, and an offline correction, using the scintillator panels as a reference, needs to be performed.
To correct this, the mean of each ToT value was calculated and subtracted from the sampled hit time.
The bottom plot of Figure~\ref{fig:timewalk} shows the corrected time difference.
During the test run, the setting for the readout speed of the second timestamp was not tuned for each individual pixel chip.
This leads to an imperfect sampling of the second timestamp, which can be seen in the top of Figure~\ref{fig:timewalk} for hits on the lower left part of the distribution ($t_{pixel}-t_{trigger}<0$ and ToT$<200$).
There, the ToT was not calculated correctly and the variance of the correction for these ToT values (see bottom of Figure~\ref{fig:timewalk}) is higher.
In contrast to the inner vertex detector, the scintillating fibre detector was not configured to send the second timestamp.
Therefore, a time walk correction was not possible for this detector.

In Figure~\ref{fig:coincidences}, the difference between the hit time of the inner vertex detector after correcting the time walk and the scintillating fibre detector is shown.
Since all hits were correlated with each other, an unintended triangular-shaped background was created.
Therefore, the used Gaussian fit function was extended by $f_{bg}(x) = c(d - |x|)$.
The performed fit showed a total time resolution of \SI{103.7\pm3.3}{ns}.

The systematic errors are dominated by the pure timing resolution of the vertex detector caused by the untuned chips, which is in the order of \SI{100}{ns}.
The sampling of the scintillator panels was done with \SI{8}{ns} on the FEB, which is the main source of systematic uncertainty for this detector.
The scintillating timing detector should have a smaller uncertainty on its timing resolution~\cite{fibre} than the systematics of the scintillator panels.
Nevertheless, despite all these limitations on the detector side, the presented results prove that the DAQ system is able to process different detector data, perform online time synchronisation to provide the required hit data to study time resolutions between the different detectors and track reconstruction.

\section{Summary and Outlook}
The paper presented detailed insights in the developed firmware parts of the Mu3e DAQ system, which are used to process the data from the different detectors to the filter farm.
Especially, the time alignment firmware used on the SWB to synchronise the different detector systems was shown.
Furthermore, a detailed description of the Mu3e Cosmic Run 2022 and first timing and tracking studies of the different detectors using a vertical slice of the Mu3e DAQ system are given.

Showing first physical plausible results for the detected cosmic rays and being able to have a consistent readout chain are first steps in commissioning the final DAQ system of the Mu3e experiment.

During the end of 2022, the commissioning of the filter farm will start.
The final detector will be constructed in the beginning of 2023 and first data taking is planned for the end of 2023.

\ifCLASSOPTIONcaptionsoff
  \newpage
\fi

\bibliographystyle{IEEEtran}

\begin{thebibliography}{10}
\providecommand{\url}[1]{#1}
\csname url@samestyle\endcsname
\providecommand{\newblock}{\relax}
\providecommand{\bibinfo}[2]{#2}
\providecommand{\BIBentrySTDinterwordspacing}{\spaceskip=0pt\relax}
\providecommand{\BIBentryALTinterwordstretchfactor}{4}
\providecommand{\BIBentryALTinterwordspacing}{\spaceskip=\fontdimen2\font plus
\BIBentryALTinterwordstretchfactor\fontdimen3\font minus
  \fontdimen4\font\relax}
\providecommand{\BIBforeignlanguage}[2]{{%
\expandafter\ifx\csname l@#1\endcsname\relax
\typeout{** WARNING: IEEEtran.bst: No hyphenation pattern has been}%
\typeout{** loaded for the language `#1'. Using the pattern for}%
\typeout{** the default language instead.}%
\else
\language=\csname l@#1\endcsname
\fi
#2}}
\providecommand{\BIBdecl}{\relax}
\BIBdecl

\bibitem{ARNDT2021165679}
K.~Arndt \emph{et~al.}, ``{Technical design of the phase I Mu3e experiment},''
  \emph{Nucl. Instr. Meth. A}, vol. 1014, p. 165679, 2021.

\bibitem{sindrum}
U.~Bellgardt \emph{et~al.}, ``{Search for the Decay $\mu^+ \rightarrow e^+ e^+
  e^-$},'' \emph{Nucl. Phys.}, vol. B299, p.~1, 1988.

\bibitem{himb}
\BIBentryALTinterwordspacing
M.~Aiba \emph{et~al.}, ``{Science Case for the new High-Intensity Muon Beams
  HIMB at PSI},'' 2021. [Online]. Available:
  \url{https://arxiv.org/abs/2111.05788} Accessed: 2022-12-12.
\BIBentrySTDinterwordspacing

\bibitem{mupix1}
H.~Augustin \emph{et~al.}, ``{The MuPix high voltage monolithic active pixel
  sensor for the Mu3e experiment},'' \emph{JINST}, vol.~10, no.~03, p. C03044,
  2015.

\bibitem{mupix2}
------, ``{MuPix7 - A fast monolithic HV-CMOS pixel chip for Mu3e},''
  \emph{JINST}, vol.~11, no.~11, p. C11029, 2016.

\bibitem{mupix3}
------, ``{The MuPix System-on-Chip for the Mu3e Experiment},'' \emph{Nucl.
  Instrum. Meth.}, vol. A845, pp. 194--198, 2017.

\bibitem{mupix4}
------, ``{Efficiency and timing performance of the MuPix7 high-voltage
  monolithic active pixel sensor},'' \emph{Nucl. Instr. Meth. A}, vol. 902, p.
  158, 2018.

\bibitem{mupix5}
------, ``{MuPix8 --- Large area monolithic HVCMOS pixel detector for the Mu3e
  experiment},'' \emph{Nucl. Instrum. Meth. A}, vol. 936, pp. 681--683, 2019.

\bibitem{mupix6}
------, ``{Performance of the large scale HV-CMOS pixel sensor MuPix8},''
  \emph{JINST}, vol.~14, no.~10, p. C10011, 2019.

\bibitem{mupix7}
------, ``{MuPix10: First Results from the Final Design},'' \emph{Proceedings
  of VERTEX2020}, 12 2020.

\bibitem{fibre}
S.~Bravar \emph{et~al.}, ``{Scintillating fibre detector for the Mu3e
  experiment},'' \emph{JINST}, vol.~12, no.~07, p. C07011, 2017.

\bibitem{tile}
H.~Klingenmeyer \emph{et~al.}, ``{Measurements with the technical prototype for
  the Mu3e tile detector},'' \emph{Nucl. Instr. Meth.}, vol. A958, p. 162852,
  2019.

\bibitem{mutrig}
H.~Chen \emph{et~al.}, ``{MuTRiG: a mixed signal Silicon Photomultiplier
  readout ASIC with high timing resolution and gigabit data link},''
  \emph{JINST}, vol.~12, no.~01, p. C01043, 2017.

\bibitem{pcie40}
J.~P. Cachemiche, P.~Y. Duval, F.~Hachon, R.~Le~Gac, and F.~R\'ethor\'e, ``{The
  PCIe-based readout system for the LHCb experiment},'' \emph{JINST}, vol.~11,
  no.~02, p. P02013, 2016.

\bibitem{de5net}
``{DE5a-NET FPGA Development Kit User Manual},'' Terasic Inc, Tech. Rep., 2019.

\bibitem{Genesys}
\BIBentryALTinterwordspacing
Digilent, ``{The Digilent Genesys 2 Kintex-7 FPGA development board}.''
  [Online]. Available:
  \url{https://reference.digilentinc.com/reference/programmable-logic/genesys-2/}
  Accessed: 2022-12-12.
\BIBentrySTDinterwordspacing

\bibitem{daq}
H.~Augustin \emph{et~al.}, ``The mu3e data acquisition,'' \emph{{IEEE
  Transactions on Nuclear Science}}, vol.~68, pp. 1833--1840, 10 2020.

\bibitem{atlasdaq}
W.~P. Vazquez \emph{et~al.}, ``{The ATLAS Data Acquisition System: from Run 1
  to Run 2},'' \emph{Nuclear and Particle Physics Proceedings}, pp. 273--275,
  2016.

\bibitem{cmsdaq}
\BIBentryALTinterwordspacing
{CMS collaboration}, ``{CMS Technical Design Report for the Level-1 Trigger
  Upgrade},'' 2013. [Online]. Available:
  \url{http://cds.cern.ch/record/1556311} Accessed: 2022-12-12.
\BIBentrySTDinterwordspacing

\bibitem{lhcbDaq}
R.~Aaij \emph{et~al.}, ``{Allen: A High-Level Trigger on GPUs for LHCb},''
  \emph{Computing and Software for Big Scienc}, vol.~4, 2020.

\bibitem{alice}
{ALICE collaboration}, ``{Real-time data processing in the ALICE High Level
  Trigger at the LHC},'' \emph{Computer Physics Communications}, vol. 242, pp.
  25--48, 2019.

\bibitem{8b10b}
A.~X. Widmer and P.~A. Franaszek, ``{A DC-Balanced, Partitioned-Block, 8B/10B
  Transmission Code},'' \emph{IBM Journal of Research and Development},
  vol.~27, p. 440, 1983.

\bibitem{bruch}
D.~vom Bruch, ``{Pixel Sensor Evaluation and Online Event Selection for the
  Mu3e Experiment},'' Ph.D. dissertation, Heidelberg University, 2017.

\bibitem{valentin}
\BIBentryALTinterwordspacing
V.~Henkys, B.~Schmidt, and N.~Berger, ``{Online Event Selection for Mu3e using
  GPUs},'' 2022. [Online]. Available: \url{https://arxiv.org/abs/2206.11535}
  {Accessed}: 2022-12-12.
\BIBentrySTDinterwordspacing

\bibitem{midas1}
S.~Ritt, P.~Amaudruz, and K.~Olchanski, ``{The MIDAS data acquisition
  system},'' \emph{Proc. IEEE 10th Real Time Conf.}, pp. 309--312, 1997.

\bibitem{midas2}
\BIBentryALTinterwordspacing
------, ``{Maximum Integration Data Acquisition System},'' 2001. [Online].
  Available: \url{https://midas.triumf.ca/} Accessed: 2022-12-12.
\BIBentrySTDinterwordspacing

\bibitem{root}
R.~Brun and F.~Rademakers, ``{ROOT - An Object Oriented Data Analysis
  Framework},'' \emph{Nucl. Inst. \& Meth. in Phys. Res. A}, vol. 389, pp.
  81--86, 1997.

\bibitem{trirec}
N.~Berger \emph{et~al.}, ``{A New Three-Dimensional Track Fit with Multiple
  Scattering},'' \emph{Nucl. Instr. Meth A.}, vol. 844, p. 135–140, 2017.

\bibitem{intrun2021}
M.~K\"oppel, ``{Mu3e Integration Run 2021},'' \emph{Proc. PoS NuFact2021}, p.
  233, 2022.

\bibitem{thomas}
T.~Rudzki, ``{The Mu3e vertex detector - construction, cooling, and first
  prototype operation},'' Ph.D. dissertation, Heidelberg University, 2022.

\bibitem{manalyzer}
\BIBentryALTinterwordspacing
K.~Olchanski, ``{MIDAS analyzer}.'' [Online]. Available:
  \url{https://bitbucket.org/tmidas/manalyzer/} Accessed: 2022-12-12.
\BIBentrySTDinterwordspacing

\end{thebibliography}

\end{document}